\documentclass[runningheads]{llncs}
\usepackage[T1]{fontenc}

\usepackage{comment}
\usepackage{cite}
\usepackage{amsmath,amssymb,amsfonts}
\usepackage{algorithmic}
\usepackage{graphicx}
\usepackage{textcomp}
\usepackage{xcolor}
\usepackage{multirow}
\usepackage{subcaption}
\usepackage{makecell} 
\usepackage{array}
\usepackage{url}
\usepackage{tabularx}
\usepackage{booktabs}

\newif\ifremark
\long\def\remark#1{
\ifremark%
        \begingroup%
        \dimen0=\columnwidth
        \advance\dimen0 by -1in%
        \setbox0=\hbox{\parbox[b]{\dimen0}{\protect\em #1}}
        \dimen1=\ht0\advance\dimen1 by 2pt%
        \dimen2=\dp0\advance\dimen2 by 2pt%
        \vskip 0.25pt%
        \hbox to \columnwidth{%
                \vrule height\dimen1 width 3pt depth\dimen2%
                \hss\copy0\hss%
                \vrule height\dimen1 width 3pt depth\dimen2%
        }%
        \endgroup%
\fi}

\remarktrue
\remarkfalse

\begin{document}

\title{TailBench++: Flexible Multi-Client, Multi-Server Benchmarking for Latency-Critical Workloads}

%




\author{Zhilin Li \and Lucia Pons \and Salvador Petit \and Julio Sahuquillo 
\and Julio Pons}

\institute{Department of Computer Engineering, 
Universitat Polit\`ecnica de Val\`encia, Spain\\
\email{lupones@disca.upv.es}
}

\maketitle

\begin{abstract}

Cloud systems have rapidly expanded worldwide in the last decade, shifting computational tasks to cloud servers where clients submit their requests. Among cloud workloads, latency-critical applications --characterized by high-percentile response times-- have gained special interest. These applications are present in modern services, representing an important fraction of cloud workloads. 
This work analyzes common cloud benchmarking suites and identifies TailBench as the most suitable to assess cloud performance with latency-critical workloads.
Unfortunately, this suite presents key limitations, especially in multi-server scenarios or environments with variable client arrival patterns and fluctuating loads.
To address these limitations, we propose TailBench++, an enhanced benchmark suite that extends TailBench to enable cloud evaluation studies to be performed in dynamic multi-client, multi-server environments. It allows reproducing experiments with varying client arrival times, dynamic query per second (QPS) fluctuations, and multiple servers handling requests. 
Case studies show that TailBench++ enables more realistic evaluations by capturing a wider range of real-world scenarios.


\end{abstract}

\section{Introduction}

Cloud systems have rapidly expanded worldwide, consisting of multiple nodes (e.g., CPUs, GPUs, and NN accelerators) interconnected among them to execute the client workloads. 
These systems typically follow a client-server model, handling 
\emph{queries per second} (QPS) rates ranging from just a few to thousands of QPS.
Unlike high-performance systems, where throughput (quantified as instructions per cycle or IPC) is the main performance metric, cloud performance is primarily evaluated with the end-to-end latency. 
In particular, tail latency —typically ranging from 95$^{th}$ to 99$^{th}$ percentile— is critical, as even a small number of high-latency requests can significantly impact the user experience \cite{talesoftail,tailatscale,tailindatacenter}.

Due to the complexity of cloud systems, research is often conducted on experimental testbeds \cite{stratus} composed of a small set of server nodes that closely mimic the behavior of production systems.
To ensure representative results, in addition to a well-designed testbed, the benchmarks used must be representative of the behavior of
\emph{real} workload, especially for latency-critical applications, which exhibit dynamic behavior over time. 
Several benchmark suites have been proposed to evaluate cloud systems. Unfortunately, most of them \cite{spec_cloud_iaas2018, perfkit_benchmarker,mlperf,bigdatabench} primarily focus on areas other than tail latency, leaving a significant gap in latency-critical workloads for cloud research.
Only TailBench \cite{tailbench} and CloudSuite \cite{cloudsuite} provide an important subset of latency-critical applications. Our analysis shows that TailBench is particularly well-suited for tail latency research due to its diverse application domains (e.g., speech/image recognition, language translation), unified harness, and source code availability. 


Nonetheless, TailBench presents important limitations that prevent these benchmarks from being used in a wide variety of studies. 
Real-world cloud systems deploy multiple worker servers and client machines, yet TailBench lacks flexibility in reproducing these scenarios. 
In this work, we identify four major issues that need to be addressed to model more realistic multi-client and multi-server scenarios. First, a TailBench server cannot start processing requests until a predefined number of clients are connected. Second, once request processing has started, the server cannot accept new clients. Third, server execution halts when all predefined clients have finished.
Fourth, the number of requests that clients can send is limited by the server.



In this paper, we propose TailBench++, which addresses the mentioned issues. 
TailBench++ is an enhanced benchmark suite designed for dynamic multi-client, multi-server cloud studies. TailBench++ expands the applicability of latency-critical workloads without altering their behavior, enabling a wider scope and more realistic evaluation studies, which is the main contribution of this work.

\section{Analysis of Existing Benchmark Suites} \label{sec:existing-benchmark-suites}

The rapid growth of cloud computing has driven the development of multiple benchmark suites aimed at helping researchers to assess cloud workloads. 
Unfortunately, many of these suites \cite{spec_cloud_iaas2018,perfkit_benchmarker,mlperf,amplab_benchmark} overlook introducing benchmarks that evaluate tail latency,
 which is the focus of this work. 
For instance, SPEC Cloud IaaS (Infrastructure as a Service) benchmark suite \cite{spec_cloud_iaas2018} and Google PerfKit Benchmarker (PKB) \cite{perfkit_benchmarker} are designed to provide coarse-grained performance metrics for evaluating cloud systems, evaluating latency as part of such metrics. 
Other benchmark suites, such as MLPerf \cite{mlperf} and the Big Data Benchmark \cite{amplab_benchmark}, evaluate latency as a key factor in meeting real-time requirements instead of at a client request level.

This section discusses benchmark suites designed for latency-critical workloads.
In this regard, the two popular benchmark suites that provide the widest range of latency-critical workloads are CloudSuite \cite{cloudsuite}, and TailBench \cite{tailbench}. 
CloudSuite offers popular online services and analytics workloads.
Similarly, TailBench provides popular online services but exclusively targets latency-critical workloads.
Table \ref{tab:benchmark-suites} highlights the key differences between CloudSuite (the latest version available, 4.0) and TailBench. 
TailBench includes more latency-critical applications (8 vs. 5) and spans a wider range of tail latencies. Both suites feature web search and key-value store applications, but CloudSuite also includes a streaming benchmark, while TailBench includes text, image, and speech recognition applications.

\begin{table}[tb]
    \centering
    \caption{Comparison of CloudSuite and TailBench benchmark suites.} \vspace{0.2cm}
    \label{tab:benchmark-suites}
    \resizebox{\textwidth}{!}{%
    \begin{tabular}{|c|c|c|c|c|c|c|c|}
    \hline
        \multirow{2}{*}{\textbf{Benchmark Suite}} 
        & \multirow{2}{*}{\textbf{Year}} 
        & \multicolumn{3}{c|}{\textbf{Latency-critical applications}} 
        & \multicolumn{3}{c|}{\textbf{Methodology}} \\ \cline{3-8}
        & 
        & \textbf{Num.} 
        & \textbf{Domains} 
        & \textbf{Range} 
        & \textbf{Source Code} 
        & \textbf{Harness} 
        & \textbf{Multi-Server} \\ \hline
        CloudSuite (4.0) \cite{cloudsuite} 
        & 2016 
        & 5 
        & \makecell[l]{Web Search and serving \\ Key-Value Stores \\ Streaming} 
        & \makecell[c]{Short - medium \\ 1ms–100ms }
        & {\LARGE $\times$}, Docker 
        & {\LARGE $\times$} 
        & {\color{green}{\LARGE \checkmark}} \\ \hline
        TailBench \cite{tailbench} 
        & 2016 
        & 8 
        & \makecell[l]{Web Search \\ Key-Value Stores \\ Transactional Databases \\ Text/Image/Speech Processing} 
        &  \makecell[c]{Very short - large \\ 10$\mu$s–10s }
        & {\color{green}{\LARGE \checkmark}} 
        & {\color{green}{\LARGE \checkmark}} 
        & {\LARGE $\times$} 
        \\ \hline
    \end{tabular}%
    }
\end{table}

\label{sec:tailbench++}
\begin{figure}[tb]
    \centering
    \includegraphics[width=0.8\linewidth]{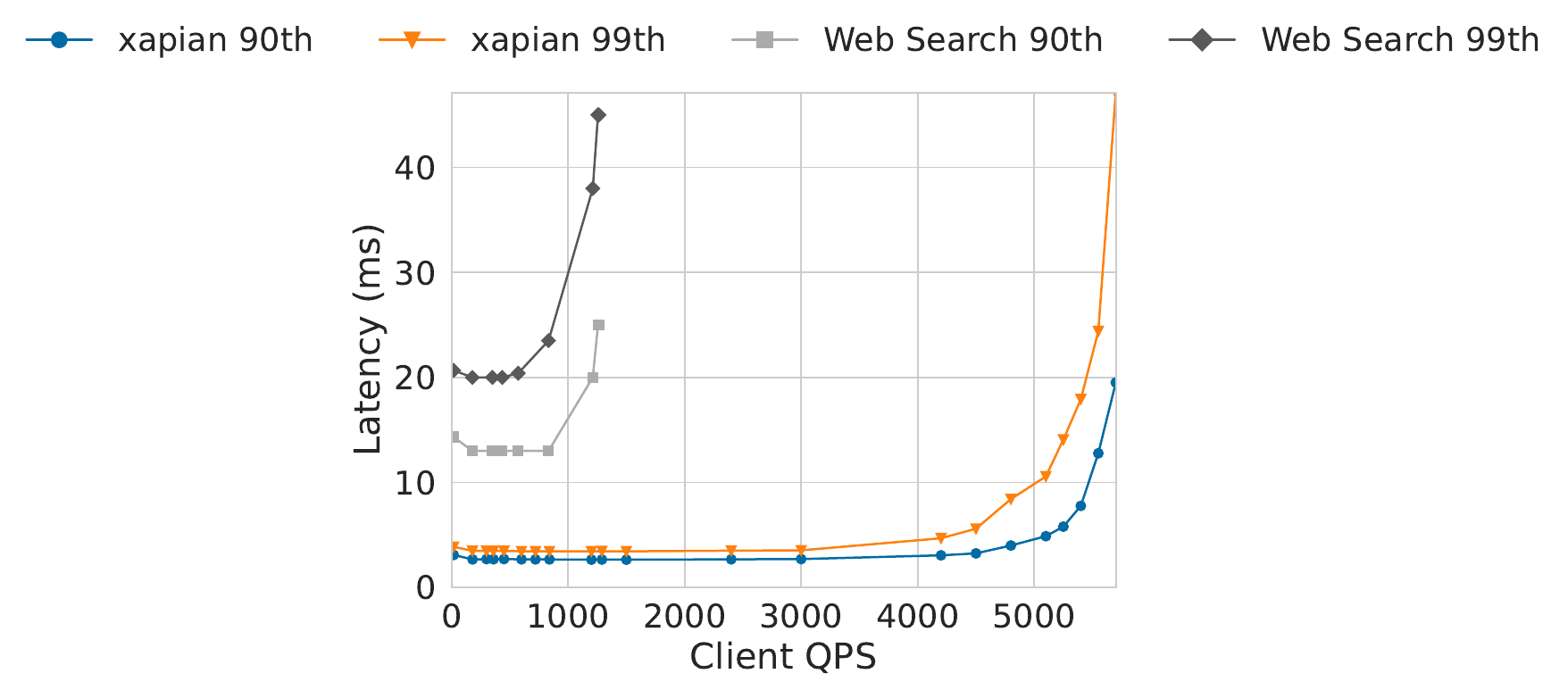}
    \vspace{-0.2cm}
    \caption{Comparison of \texttt{Web Search} (CloudSuite) and \texttt{xapian} (TailBench).}
    \label{fig:websearchvsxapian}
\end{figure}

To further illustrate the differences, a direct comparison is made between the web search domain applications --\texttt{Web Search} and \texttt{xapian}-- from CloudSuite and TailBench, respectively.
\texttt{xapian} uses an index of Wikipedia from 2013, and 
\texttt{Web Search} is set to use the pre-generated Solr\cite{apache_solr} index, which is similar in size (14GB) to the TailBench one (15GB). For both applications, we launched three client processes that connect to one server process. 
Figure \ref{fig:websearchvsxapian} shows the 
latency obtained for both workloads as the client QPS increases\footnote{See Section \ref{sec:experimental-testbed} for details on the experimental configuration.}. 
As observed, \texttt{xapian} exhibits a broader tail latency range, starting below 5ms, whereas \texttt{Web Search} remains above 10ms. 
As \texttt{Web Search} lacks direct query rate control and requires indirect tuning, it results in a shorter load span. 
Furthermore, performance degradation occurs earlier in CloudSuite (after $QPS=1000$) compared to TailBench (after $QPS=4000$), indicating better scalability in the version of the web search application of TailBench.

Beyond applications, methodologies differ significantly. TailBench provides full source code, allowing in-depth analysis and modifications, whereas CloudSuite relies on Docker-based deployment, simplifying setup but limiting flexibility. TailBench also provides a unified execution harness, simplifying benchmark execution and evaluation, 
 while CloudSuite’s workloads have separate interfaces, which complicates configuration and result interpretation. 
Finally,  CloudSuite applications support multi-server configurations, while TailBench is limited to a single server per experiment.

{\bf To take away:} \textit{
CloudSuite is well-suited for distributed multi-server workloads with easy deployment but has a narrow tail latency range and limited code availability. In contrast, TailBench is more developer-friendly, offering broader LC application coverage and better support for tail latency.
}

\section{Motivation} \label{sec:motivation}

After analyzing existing benchmark suites, we concluded that TailBench is the most suitable for research focused on tail latency.
This suite includes eight representative latency-critical applications. To make this paper self-contained, we briefly describe each benchmark:

\begin{itemize}

\item \texttt{img-dnn}: Handwriting recognition using OpenCV with random samples from the MNIST dataset.

\item \texttt{masstree}: Fast in-memory key-value store with low latency demands due to multiple operations per user request.

\item \texttt{moses}: Statistical machine translation system processing dialogue segments from the English-Spanish corpus of opensubtitles.org.

\item \texttt{shore}: Disk-based transactional database using the TPC-C benchmark, with data and logs stored on an SSD.

\item \texttt{silo}: In-memory transactional database optimized for multicore systems, using TPC-C with different storage/access methods.

\item \texttt{specjbb}: Java middleware benchmark for business service applications with strict latency guarantees.

\item \texttt{sphinx}: Computationally intensive speech recognition system, vital for speech-driven interfaces like Siri, Google Now, and IBM Speech to Text.

\item \texttt{xapian}: C++-based search engine used in software frameworks (e.g., Catalyst) and websites (e.g., Debian wiki).

\end{itemize}

Unfortunately, TailBench has key limitations that prevent this suite from being used in more realistic, large-scale, multi-server scenarios with dynamic client behavior. These issues primarily arise from the server configuration being too restrictive. 
In this work, we identify and address four main limitations of TailBench:

\begin{enumerate}
    \item The server must wait for a fixed number of clients to connect before starting to process requests. 
    \item New client connections are not accepted once the server begins processing.
    \item If all clients disconnect, the server terminates.
    \item The total number of requests is predetermined in the server configuration, and the experiment ends when this target is reached.
\end{enumerate}

These limitations mean that TailBench applications cannot be used in many realistic scenarios representative of modern computing services, where client numbers and request rates fluctuate over time, sometimes dropping to zero (e.g., interactive workloads with diurnal patterns \cite{atikoglu2012workload,resource-central}). 

These insights motivated us to implement an extended version of the TailBench, aimed at opening new research scopes. 
It is worth noting that the goal of the proposed benchmark suite is to increase their applicability and broaden the range of potential use cases without altering the behavior of the applications.

\section{Taibench++'s Features}
\label{sec:tailbench++}

\begin{figure}[tb]
    \centering
    \includegraphics[width=1.0\linewidth]{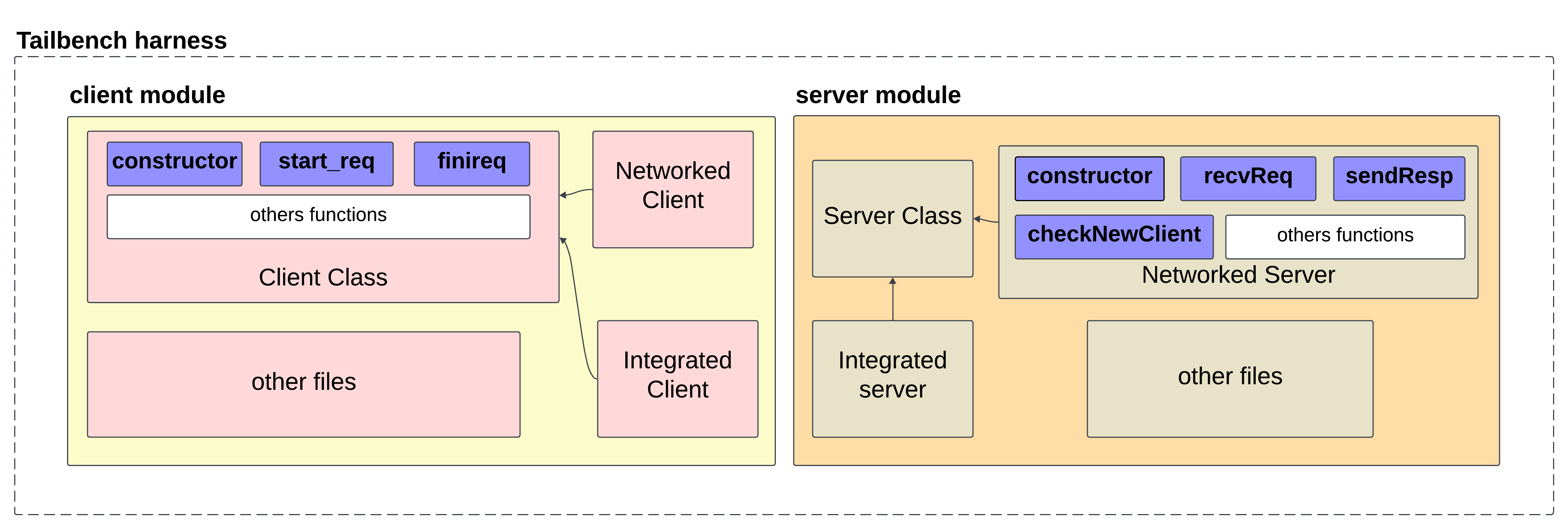}
    \caption{Overview of TailBench++ harness. Components where new features have been introduced are highlighted in blue.}
    \label{fig:harness}
\end{figure}



TailBench++ extends the original TailBench suite to support multi-server environments, dynamic load variations, and an unconstrained number of clients. 
As TailBench provides a harness that controls application execution, load generation, and statistics collection, modifications have been made to this component. TailBench offers two configurations: Networked and Integrated. In the Networked configuration, the client and server run in separate programs (on the same or different machines).
Whereas in the Integrated configuration, the client and server run in the same process; thus, it is not suitable for multi-client and multi-server scenarios. Therefore, all modifications focus on the Networked configuration.

Figure \ref{fig:harness} shows a block diagram with the components of the harness, which are grouped into two main modules: the client and the server. The seven components of the harness that have been modified are highlighted in blue.
Below, we discuss the implemented extensions.

\vspace{0.2 cm}
\noindent {\it Feature 1. Unconstrained number of clients.}
\vspace{0.1 cm}

TailBench's server previously waited for a predefined number of clients to connect before processing requests (defined in the \texttt{constructor} of the server module). 
In contrast, TailBench++ allows the server to accept new client connections dynamically, thus removing this limitation.
The server constructor is no longer responsible for accepting new client connections, allowing it to start running without having to wait for a predefined number of clients. It is now the function \texttt{recvReq} responsible for accepting new client connections by using a newly added function, \texttt{checkNewClient}, which monitors the arrival of new connections.



\vspace{0.2 cm}
\noindent {\it Feature 2. Persistent server.}
\vspace{0.1 cm}

Real-world applications require continuous server availability, which cannot be reproduced with TailBench as the server terminates when the predefined clients disconnect.
TailBench++ ensures persistence by keeping the server idle and monitoring for new client connections, making it agnostic to the number of clients.
This feature was implemented alongside \textit{Feature 1. Unconstrained number of clients}, as both required modifying the same component of the server module. In TailBench, the \texttt{recvReq} function checks for connected clients and terminates the server when none remain. In contrast, TailBench++ allows the server to stay alive and monitor new clients using \texttt{checkNewClient} function.



\vspace{0.2 cm}
\noindent {\it Feature 3. Independent client behavior.}
\vspace{0.1 cm}

In a cloud environment, clients operate independently, varying in request volume, rate, and timing. This independent client behavior cannot be reproduced in TailBench as it forces all clients to send the same number of requests, which is defined on the server side.
This behavior has been changed in TailBench++ so each client has its own workload, better reproducing real-world scenarios like users selecting different streaming content.
To implement this feature, changes were required both in the client and server modules. In TailBench, the \texttt{sendResp} function from the server module sets the request limit. In TailBench++, this control is shifted to the client module: the client constructor now defines the total number of requests at initialization, and the \texttt{finireq} function, which tracks the response times for each request, has been extended to monitor the total number of queries made and terminate the client upon reaching this limit.


\vspace{0.2 cm}
\noindent {\it Feature 4. Variable client load.}
\vspace{0.1 cm}

User load is constantly changing as demand may fluctuate based on factors such as time, content, or external events. For example, on a streaming platform, a user may initially watch a few episodes in a row and later switch to sporadic viewing. 
In TailBench, however, it is not possible to reproduce such a scenario as it enforces a fixed client request rate. 
To overcome this limitation, TailBench++ allows dynamic load variation, enabling clients to adjust request rates during execution.
The modifications were mainly made to the client module.
Optional parameters have been added to the client constructor to define load variation. In the \texttt{start\_req}, which handles request generation and tracks the time a query is sent, additional functionality was introduced to monitor and dynamically adjust the client’s load during execution.


\section{Experimental Testbed} \label{sec:experimental-testbed}

\begin{figure}[tb]
    \centering
    \includegraphics[width=0.9\linewidth]{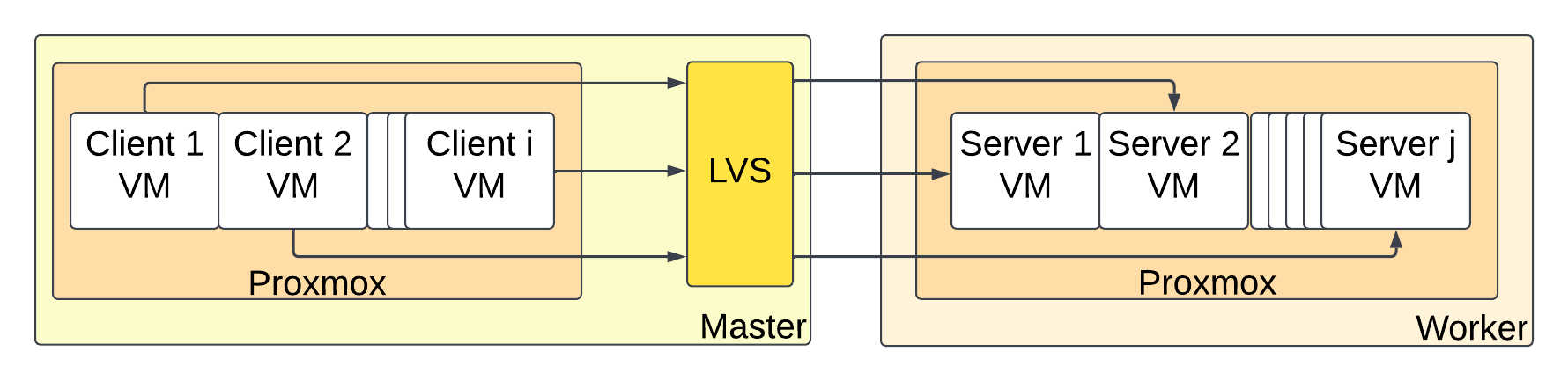}
    \caption{Data flow of client requests to servers in the experimental testbed.}
    \label{fig:system}
\end{figure}

\begin{table}[tb]
\caption{Master and worker nodes hardware details.} \vspace{0.2cm}
    \centering
    \begin{tabular}{|c|c|c|} \hline 
        \textbf{Component} & \textbf{Master node}  & \textbf{Worker node} \\\hline 
       Processor &  Intel Xeon E5-2658A v3 & 2 x Intel Xeon Gold 6438Y+\\\hline 
        L1 ICache & 32 KB & 32 KB \\\hline 
        L1 Dcache & 32 KB & 48 KB \\\hline 
        L2 Cache & 256 KB & 2MB \\\hline 
        L3 Cache & 30 MB & 60MB \\\hline
        Memory & 32 GB (1 x DDR3 1066.5 MHz)  & 256 GB (16 x 16 GB DDR5 2400 MHz) \\\hline
    \end{tabular}
    \label{tab:xpl2-xpl10}
\end{table}


\noindent{\bf System Specifications.} Experiments were conducted in a real system testbed made up of two physical machines: a master node handling TailBench clients and distributing connections via Linux Virtual Server (LVS) \cite{linuxvirtualserver}, and a worker node running the server workloads.

The master node is equipped with an Intel Xeon E5-2658A v3 processor \cite{intel_xeon_e5_2658a_v3}. This processor has 12 cores and a 30-MB LLC (Last Level Cache). Regarding the DRAM, it holds a 32GB DDR3 DIMM working at 1066.5 MHz.
The worker node is used to run the servers from TailBench workloads. It is a two-socket node equipped with two Intel Xeon Gold 6438Y+ processors\cite{intel_xeon_gold_6438y}. Each one has 32 cores and a 60-MB LLC. Its main memory provides 256 GB with 16 DDR5 DIMMs of 2400 MHz. 
Detailed information about the hardware specifications of each machine is summarized in Table \ref{tab:xpl2-xpl10}.

The master and worker nodes are interconnected via a D-Link DXS 1210-28T 24x10G Base T\cite{dlink_dxs1210_2024} switch with 10Gbps Ethernet. Both nodes run Debian Linux 12 (kernel version 6.8.4-3-pve) with Proxmox VE support.


\vspace{0.3cm}
\noindent{\bf VM Infrastructure} To create a realistic multi-client, multi-server cloud environment, we used virtualization with Proxmox VE\cite{Proxmox}, popularly employed to manage virtual machines (VMs) and containers. To evaluate the newly implemented features, clients and servers run in separate VMs. For the experiments presented in this work, a total of five VMs were deployed:
\begin{itemize}  
    \item \textbf{Servers (2 VMs, worker node)}: 4 physical cores, 16GB RAM each.  
    \item \textbf{Clients (3 VMs, master node)}: 3 physical cores, 8GB RAM each.  
\end{itemize}  

For multi-server experiments, clients send requests to LVS, an open-source load-balancing solution integrated into the Linux kernel designed to distribute network traffic across multiple servers. LVS distributes requests based on a load-balancing policy (by default, round-robin) to ensure scalability.
Figure~\ref{fig:system} illustrates the data flow under this configuration, where requests from a variable number of clients ($i$) to a variable number of servers ($j$).
In single-server setups, client VMs connect directly to the server VM, without the need for LVS.

Finally, we would like to remark that, similarly to TailBench, TailBench++ operates independently of the underlying hardware-software system. This means it is not dependent on Proxmox VE, and it can support an arbitrary number of client and server nodes, with or without VMs. Additionally, client distribution across servers is not restricted to LVS and can be managed using alternative tools such as Nginx \cite{nginx} or HAProxy \cite{haproxy}.


\section{TailBench++ Workload Characterization} \label{sec:tailbench-characterization}


\subsection{Validation of TailBench++ Application Behavior}

\begin{table}[tb]
\caption{Number of client threads generating requests in each benchmark.}
        \vspace{0.2cm}
        \centering
        \resizebox{\textwidth}{!}{%
        \begin{tabular}{|c|c|c|c|c|c|c|c|c|c|}
        \hline
        \textbf{} & \textbf{img-dnn++} & \textbf{masstree++} & \textbf{moses++} & \textbf{shore++} & \textbf{silo++} & \textbf{specjbb++} & \textbf{sphinx++} & \textbf{xapian++} \\
        \hline
        \textbf{Threads} & 12 & 4 & 2 & 2 & 8 & 8 & 1 & 2\\
        \hline
        \end{tabular}
        }    
    \label{tab:hilosClients}
\end{table}

\begin{figure*}[tb]
    \centering
    \includegraphics[trim=0 405 0 0,clip,width=0.4\textwidth]{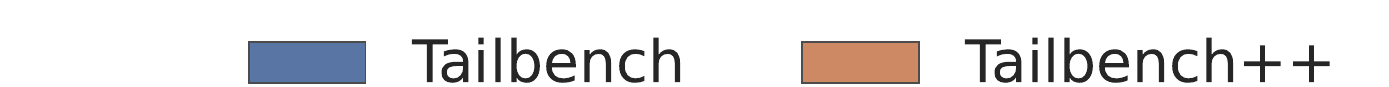}\\
      \captionsetup[subfloat]{position=bottom}
    \begin{subfigure}{0.249\linewidth}
        \includegraphics[trim=0 0 0 45,clip, width=\linewidth]{img-dnn_boxplot.pdf}
        \caption{img-dnn}
        \label{fig:imgdnn_boxplot}
    \end{subfigure}
    \hfill
    \begin{subfigure}{0.24\linewidth}
        \includegraphics[trim=36 0 0 45,clip, width=\linewidth]{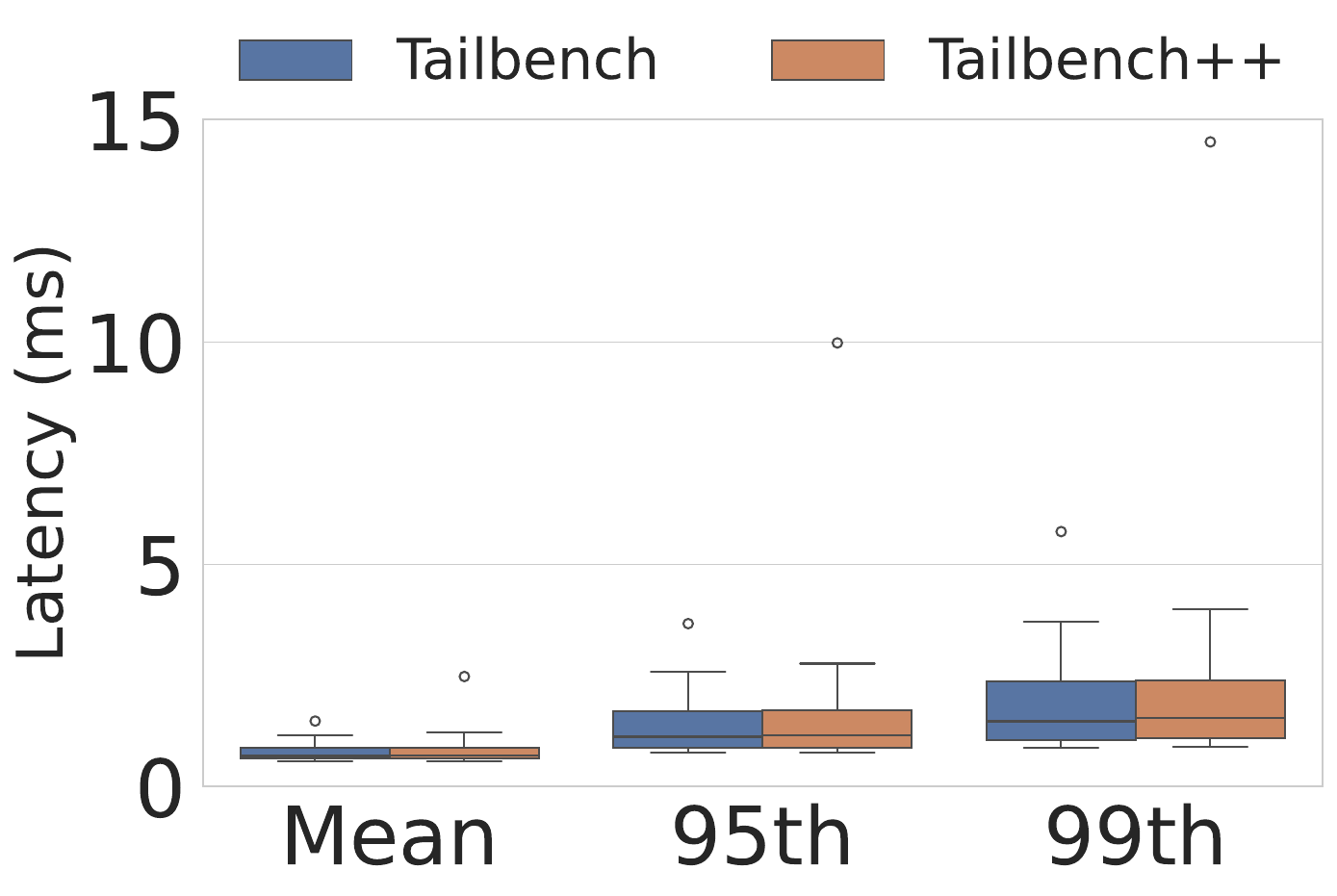}
        \caption{masstree}
        \label{fig:masstree_boxplot}
    \end{subfigure}
    \hfill
    \begin{subfigure}{0.24\linewidth}
        \includegraphics[trim=36 0 0 45,clip, width=\linewidth]{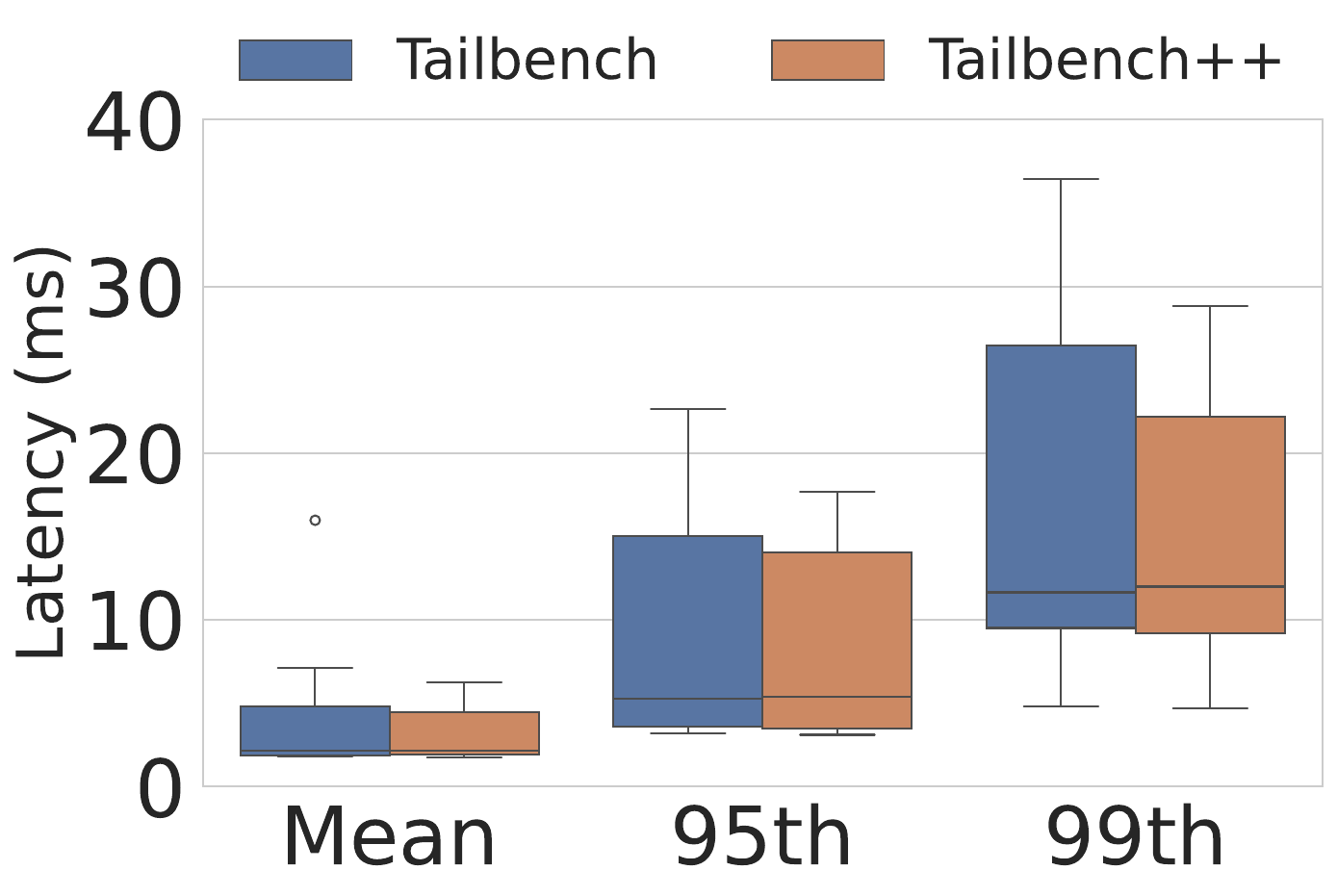}
        \caption{shore}
        \label{fig:shore_boxplot}
    \end{subfigure}
    \hfill
    \begin{subfigure}{0.235\linewidth}
        \includegraphics[trim=36 0 0 45,clip, width=\linewidth]{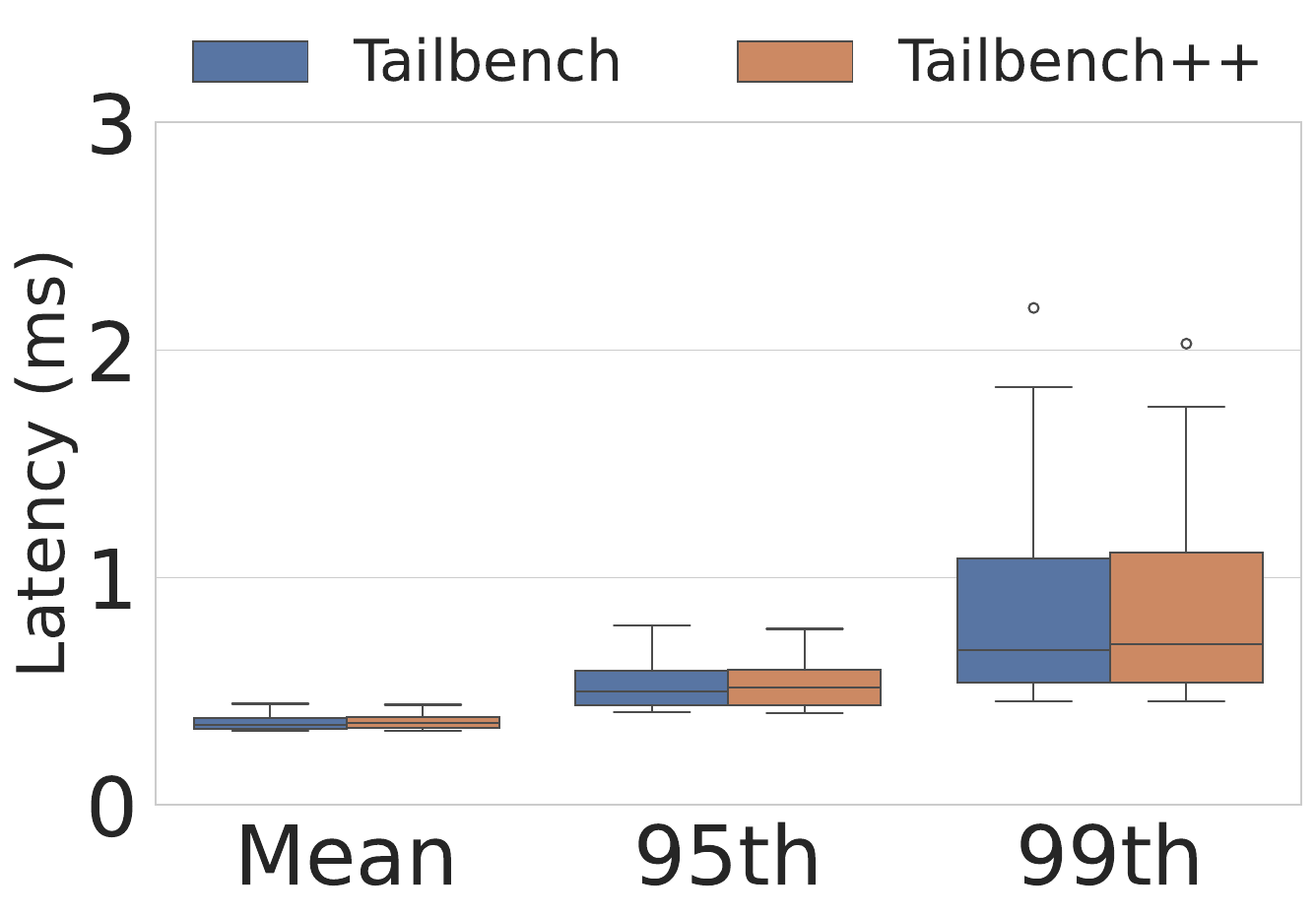}
        \caption{silo}
        \label{fig:silo_boxplot}
    \end{subfigure}
    
    \vspace{1em}
    
    \begin{subfigure}{0.242\linewidth}
        \includegraphics[trim=0 0 0 45,clip, width=\linewidth]{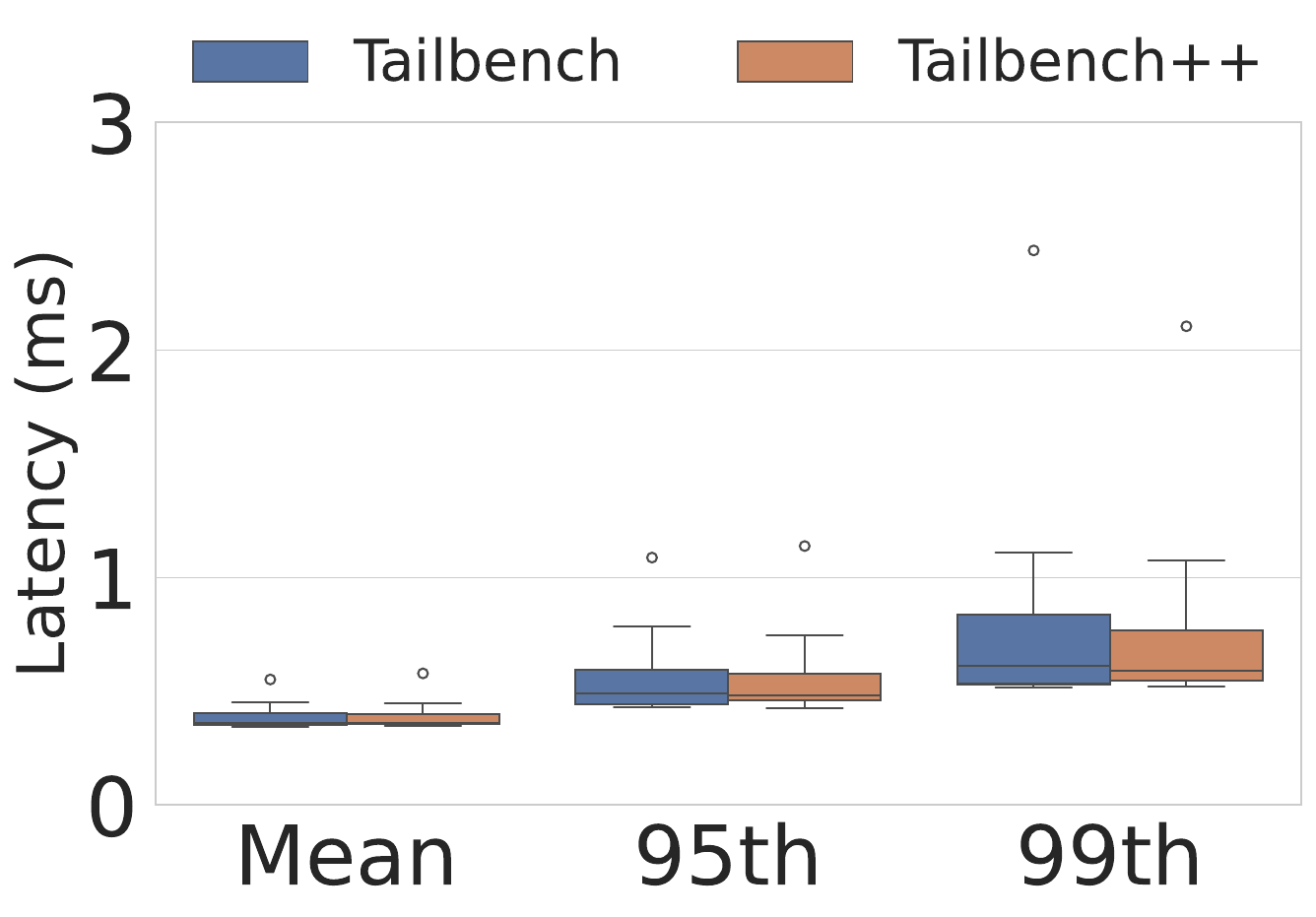}
        \caption{specjbb}
        \label{fig:specjbb_boxplot}
    \end{subfigure}
    \hfill
    \begin{subfigure}{0.255\linewidth}
        \includegraphics[trim=36 0 0 45,clip, width=\linewidth]{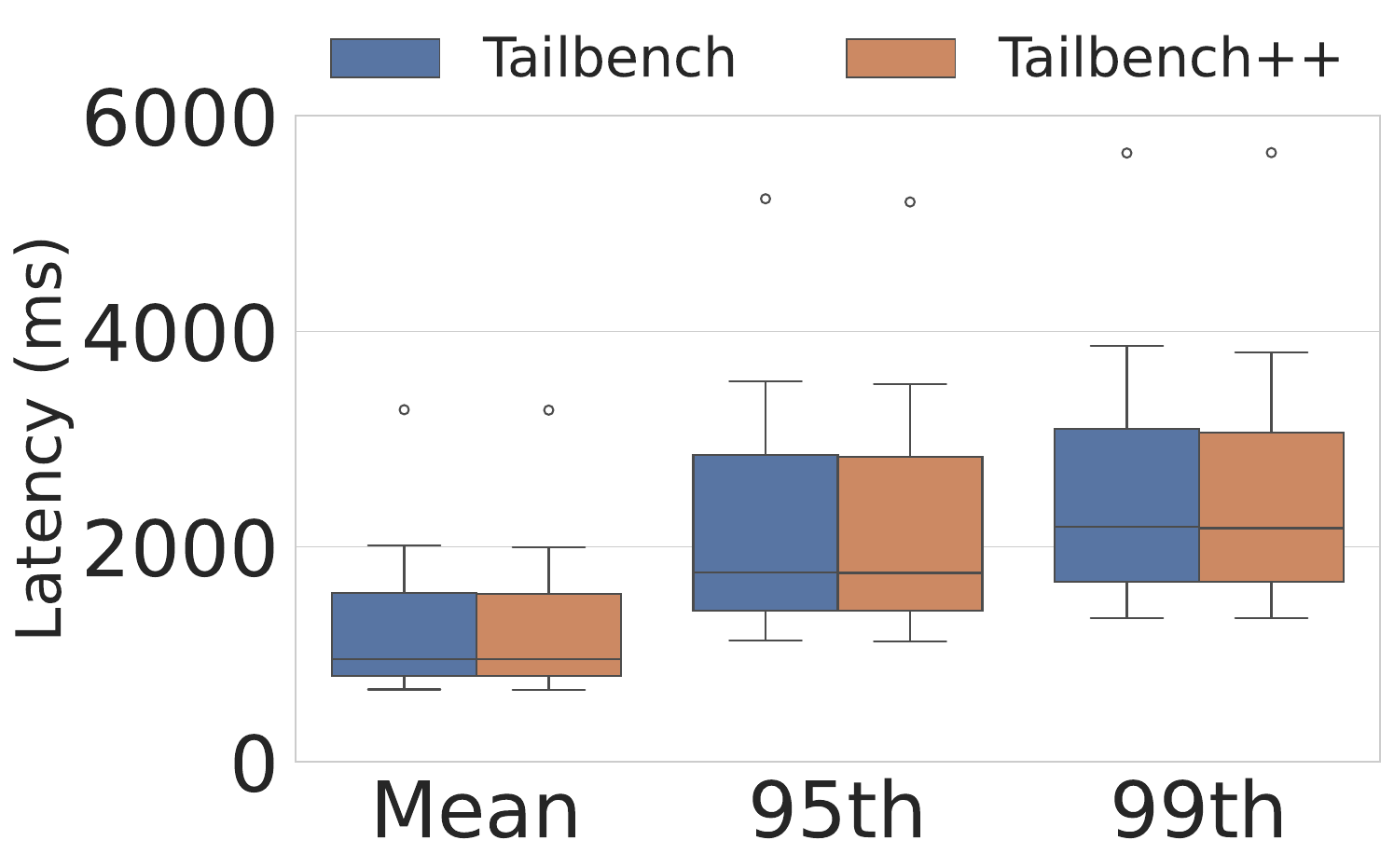}
        \caption{sphinx}
        \label{fig:sphinx_boxplot}
    \end{subfigure}
    \hfill
    \begin{subfigure}{0.238\linewidth}
        \includegraphics[trim=36 0 0 45,clip, width=\linewidth]{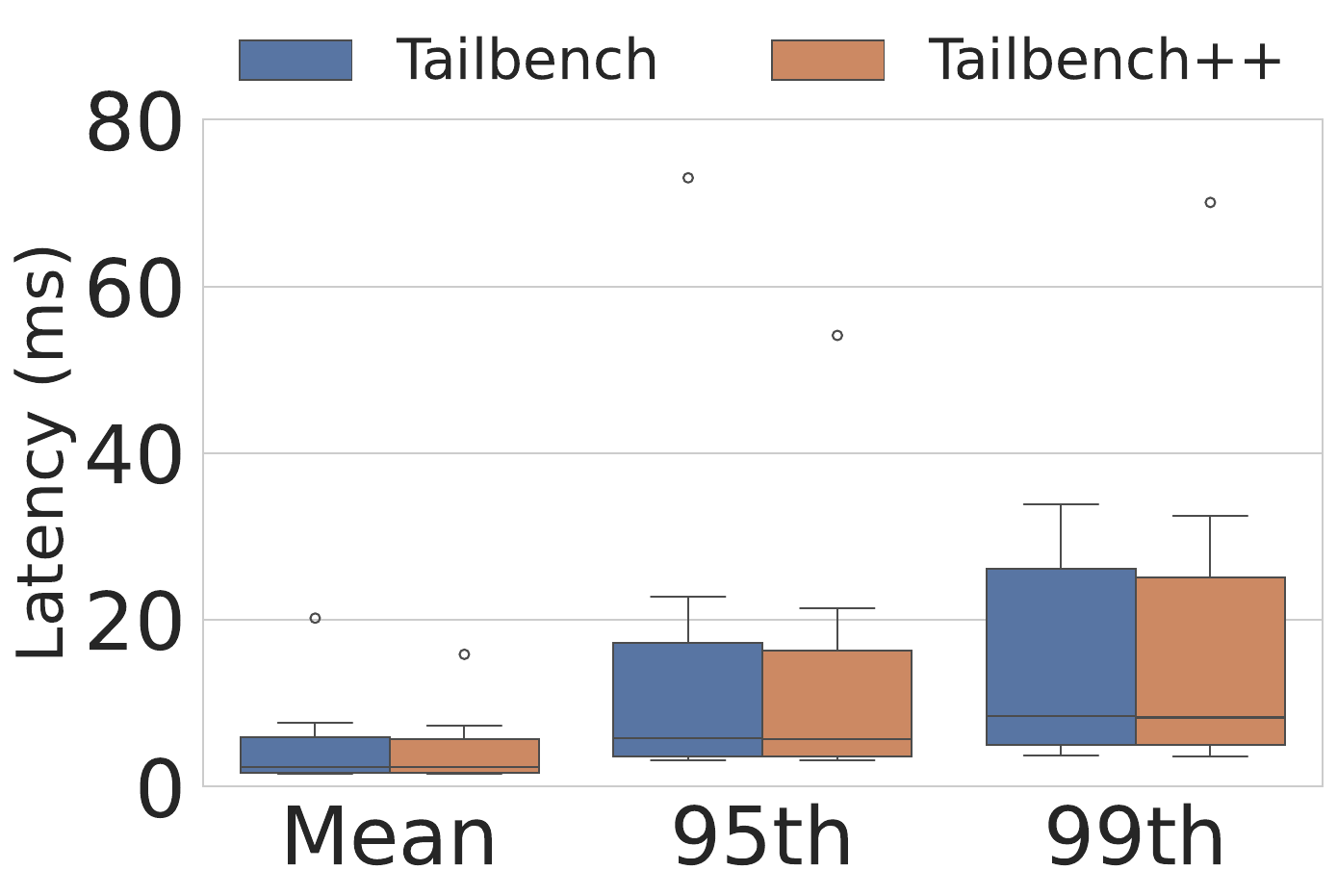}
        \caption{xapian}
        \label{fig:xapian_boxplot}
    \end{subfigure}
    \hfill
    \begin{subfigure}{0.235\linewidth}
        \includegraphics[trim=36 0 0 45,clip, width=\linewidth]{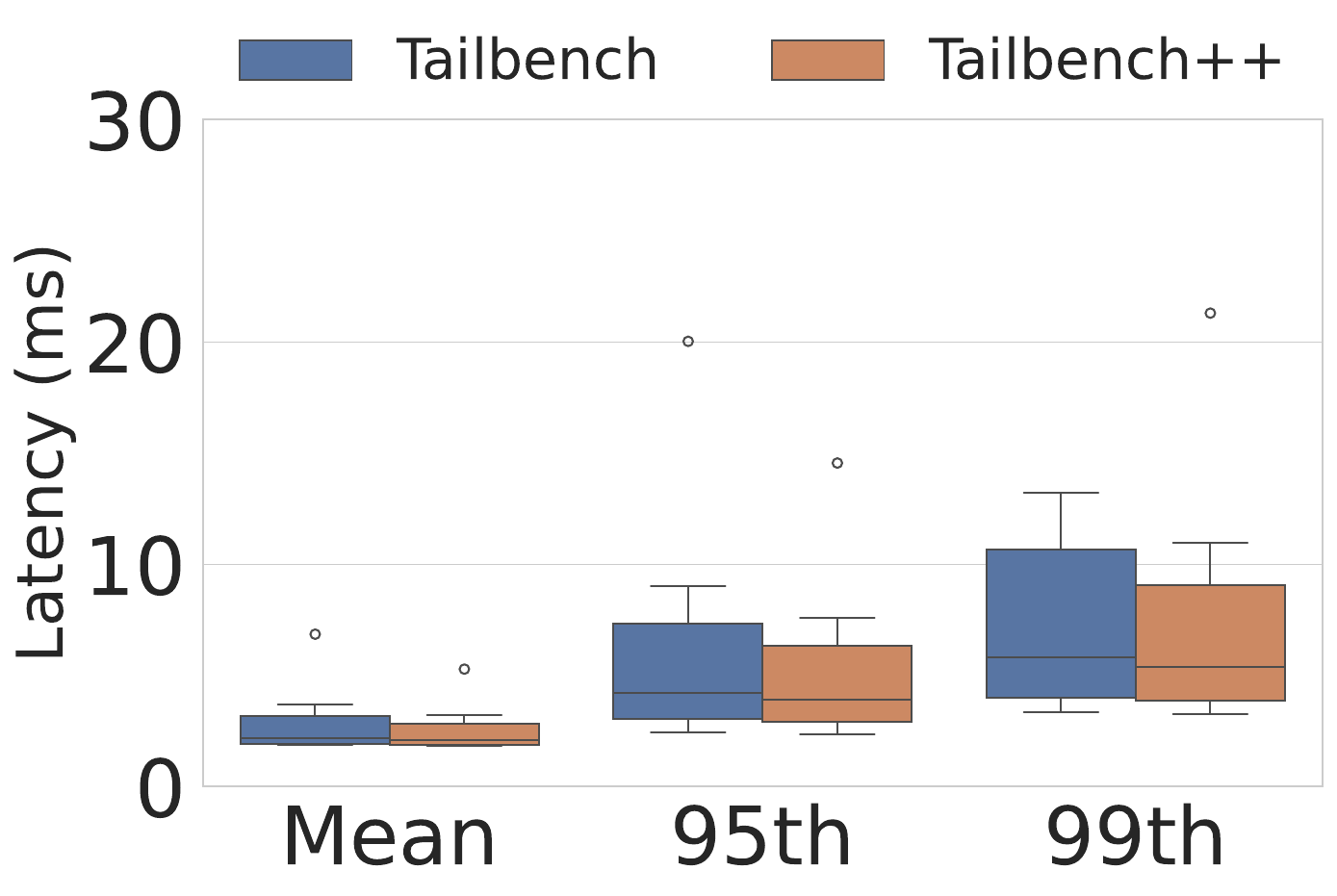}
        \caption{moses}
        \label{fig:mose_boxplot}
    \end{subfigure}

    \caption{Boxplots showing the distribution of the performance metrics (mean, 95th percentile, and 99th percentile latency in ms) for TailBench and TailBench++.}
    \label{fig:boxplots}
\end{figure*}

This section aims to prove that the behavior of applications in TailBench++ matches that of TailBench.
For a fair comparison, both suites were tested under identical conditions: a single-threaded server and a multi-threaded client. The number of client threads (see Table \ref{tab:hilosClients}) was empirically obtained to ensure sufficient load generation (QPS) while preserving the Zipfian distribution of request times, maintaining representativeness \cite{PONS2022194}.

Due to the variability that arises in experiments conducted on real machines, we obtained the latency distributions across a wide range of QPS values. Each experiment (corresponding to a specific QPS) was repeated thirteen times.
Figure~\ref{fig:boxplots}\footnote{Note that, due to Quality of Service (QoS) requirements ranging from milliseconds to seconds, different time scales have been employed to simplify the analysis.} compares the distribution of the mean, 95$^{th}$, and 99$^{th}$ percentiles latencies of applications (one graph per application) in TailBench and TailBench++. 
The boxplots for each suite exhibit nearly identical medians and interquartile ranges in most applications, 
with the only exception of \texttt{shore} and \texttt{moses} where a small deviation can be appreciated at high latencies due to variability in the disk access time.

\begin{table*}[tb]
    \centering
    \caption{\small Welch’s t-test results comparing TailBench and TailBench++ across different latency metrics. Each cell represents the T-statistic / P-value.} \vspace{0.2cm}
    \renewcommand{\arraystretch}{1.2}
     \resizebox{\textwidth}{!}{%
    \begin{tabular}{|l|c|c|c|c|c|c|c|c|}
        \hline
        \textbf{Metric} & \textbf{img-dnn} & \textbf{masstree} & \textbf{sphinx} & \textbf{silo} & \textbf{moses} & \textbf{shore} & \textbf{xapian} & \textbf{specjbb} \\
        \hline
        95$^{th}$  & 0.24 / 0.81 & -0.48 / 0.64 & 0.01 / 0.99 & 0.12 / 0.91 & 0.39 / 0.71 & -0.23 / 0.82 & 0.07 / 0.94 & -0.10 / 0.92 \\
        99$^{th}$  & 0.20 / 0.85 & -0.15 / 0.89 & 0.01 / 0.99 & 0.11 / 0.92 & 0.42 / 0.68 & -0.32 / 0.75 & 0.05 / 0.96 & -0.06 / 0.95 \\
        Mean          & 0.16 / 0.87 & -0.27 / 0.79 & 0.01 / 0.99 & 0.22 / 0.83 & 0.42 / 0.68 & -0.36 / 0.73 & 0.07 / 0.94 & -0.01 / 0.99 \\
        \hline
    \end{tabular}
    }
    
    \label{tab:t_test_results}
\end{table*}

To further prove this fact, we carried out Welch’s t-test \cite{welch-t-test} to compare the distributions of the 95th percentile latency, 99th percentile latency, and mean latency across different QPS. 
We define the following hypothesis: 
\begin{itemize}
    \item Null Hypothesis ($H_0$): There is no significant difference between the latency distributions of TailBench and TailBench++.
    \item Alternative Hypothesis ($H_1$): There is a significant difference.
\end{itemize}
The test results are presented in Table \ref{tab:t_test_results}. In all cases, the t-statistic ($|t|$) is small ($<2$) and p-value > 0.05; thus, no significant difference appears across all applications, meaning that the null hypothesis is retained. 
Therefore, we can affirm that the implementation of the new features in TailBench++ has not modified the behavior of the benchmarks, and can be considered representative.


\subsection{Multi-Server Characterization}



\begin{figure*}[tb]
    \centering
    \includegraphics[trim=0 340 0 0,clip,width=0.9\textwidth]{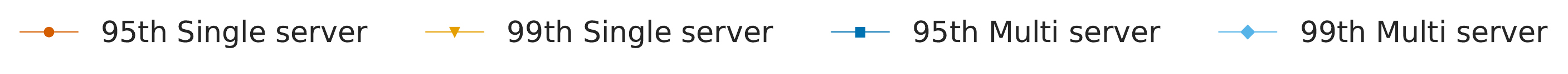}\\
    \begin{subfigure}{0.25\linewidth}
        \includegraphics[trim=0 0 0 0,clip, width=\linewidth]{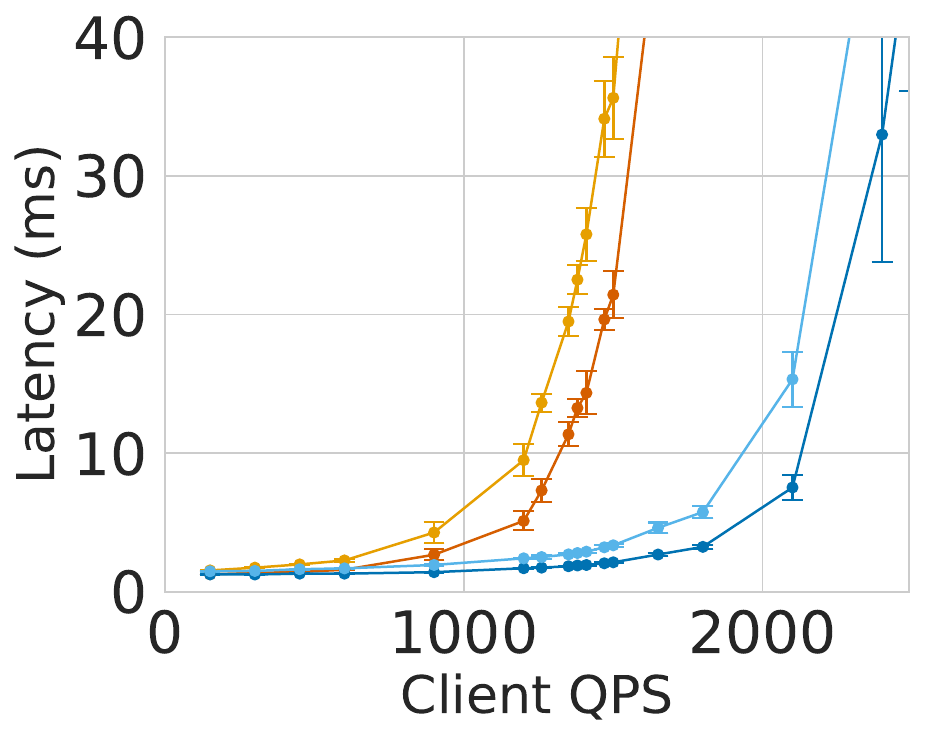}
        \caption{img-dnn++}
        \label{fig:imgdnn++lvs}
    \end{subfigure}
    \hfill
    \begin{subfigure}{0.24\linewidth}
        \includegraphics[trim=40 0 0 0,clip,width=\linewidth]{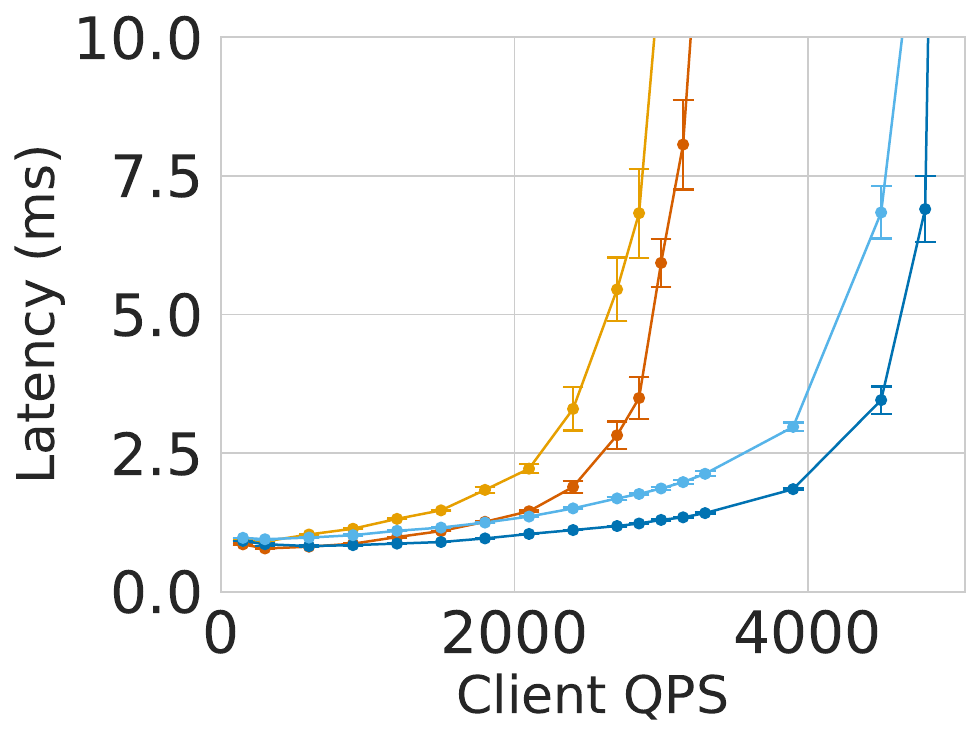}
        \caption{masstree++}
        \label{fig:masstree++lvs}
    \end{subfigure}
    \hfill
        \begin{subfigure}{0.23\linewidth}
        \includegraphics[trim=32 0 0 0,clip,width=\linewidth]{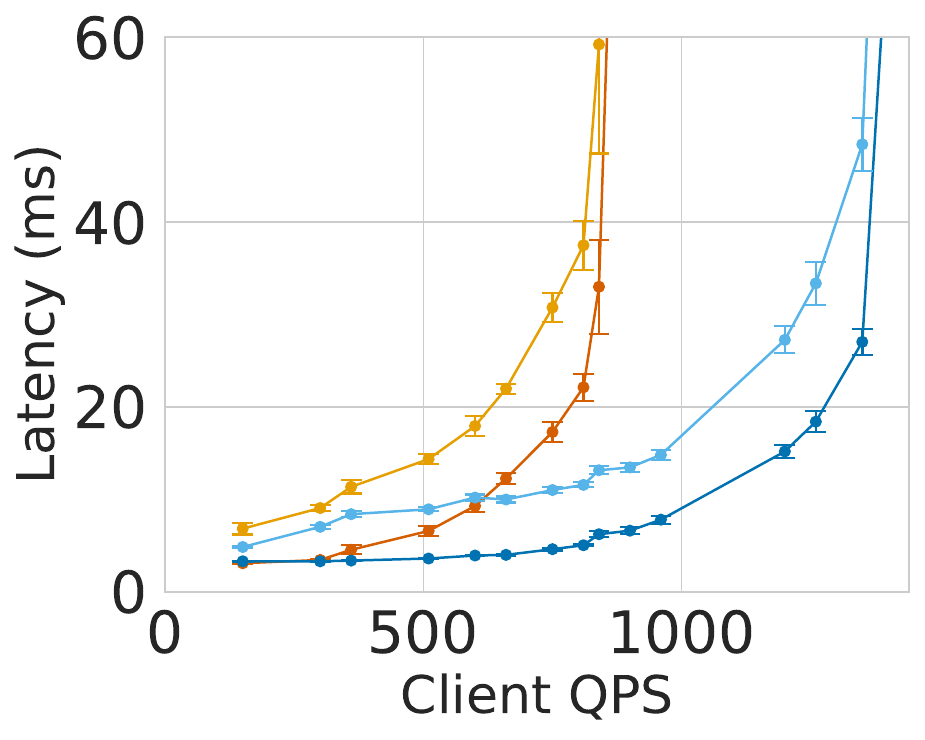}
        \caption{shore++}
        \label{fig:shore++lvs}
    \end{subfigure}
    \hfill
    \begin{subfigure}{0.237\linewidth}
        \includegraphics[trim=32 0 0 0,clip,width=\linewidth]{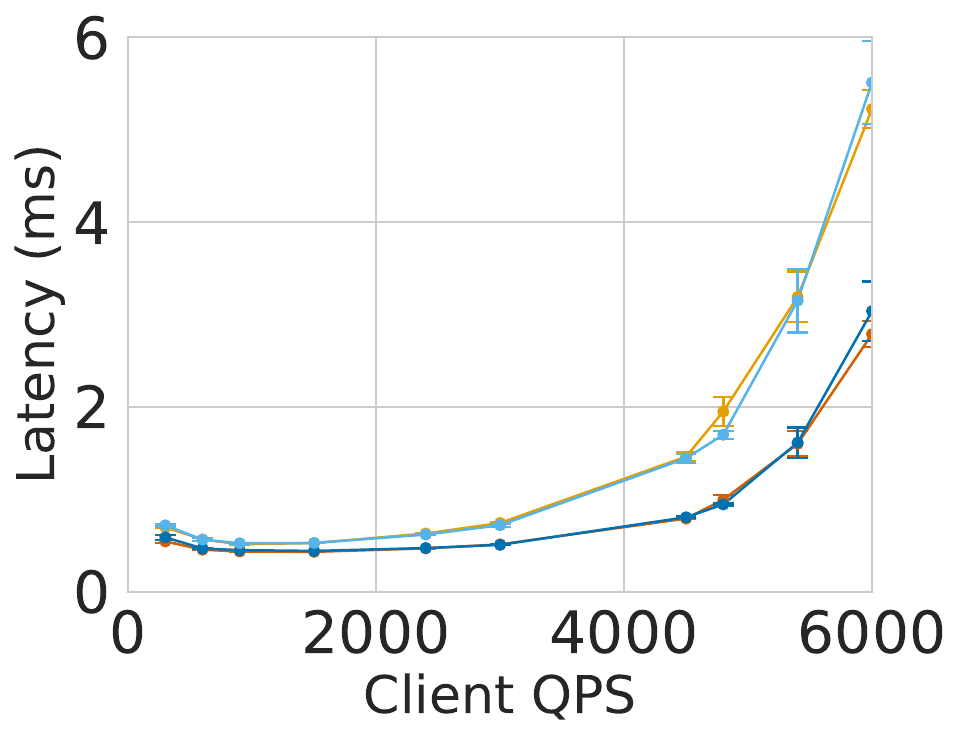}
        \caption{silo++}
        \label{fig:silo++lvs}
    \end{subfigure}
    \hfill

    \vspace{1em}
    
    \begin{subfigure}{0.247\linewidth}
        \includegraphics[trim=0 0 0 0,clip,width=\linewidth]{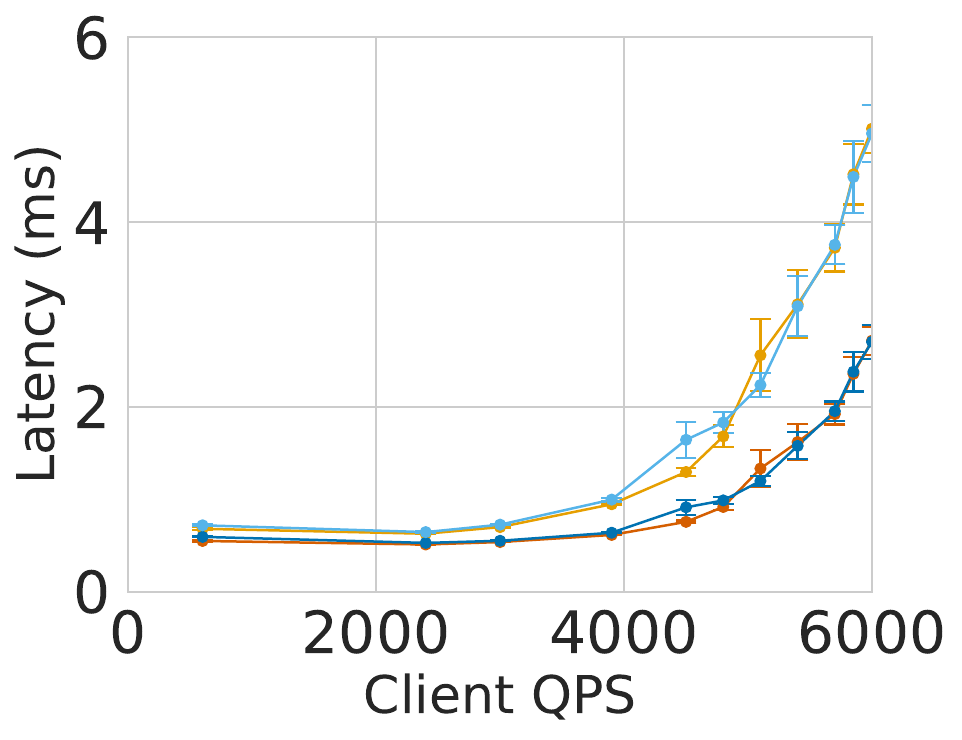}
        \caption{specjbb++}
        \label{fig:specjbb++lvs}
    \end{subfigure}
    \hfill
    \begin{subfigure}{0.247\linewidth}
        \includegraphics[trim=36 0 0 0,clip,width=\linewidth]{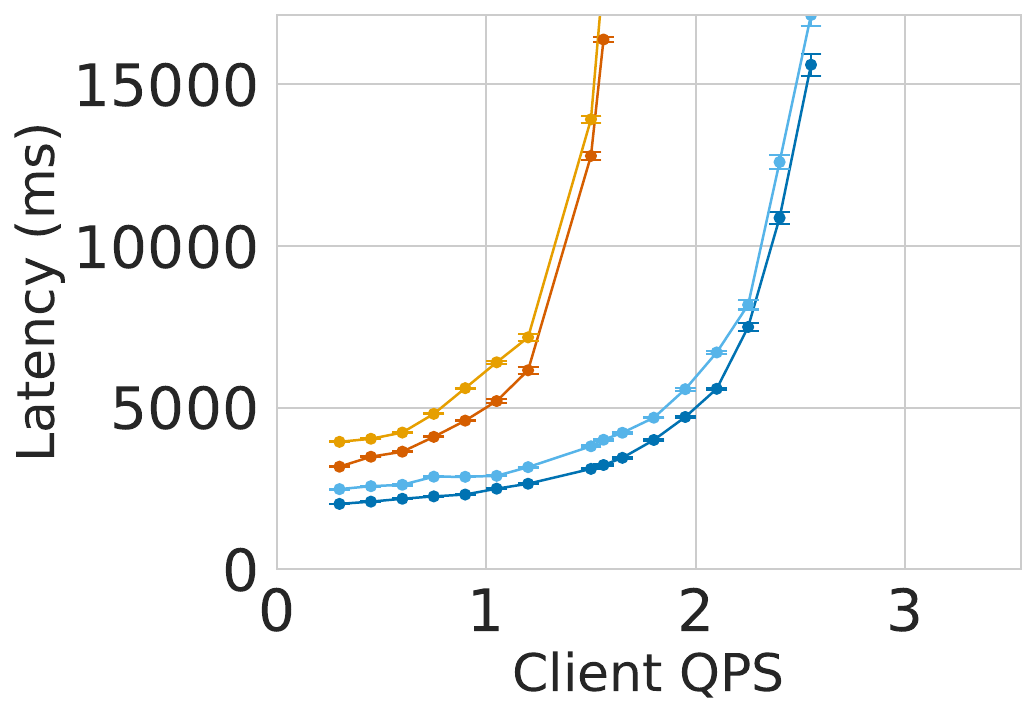}
        \caption{sphinx++}
        \label{fig:sphinx++lvs}
    \end{subfigure}
    \hfill
    \begin{subfigure}{0.22\linewidth}
        \includegraphics[trim=30 0 0 0,clip,width=\linewidth]{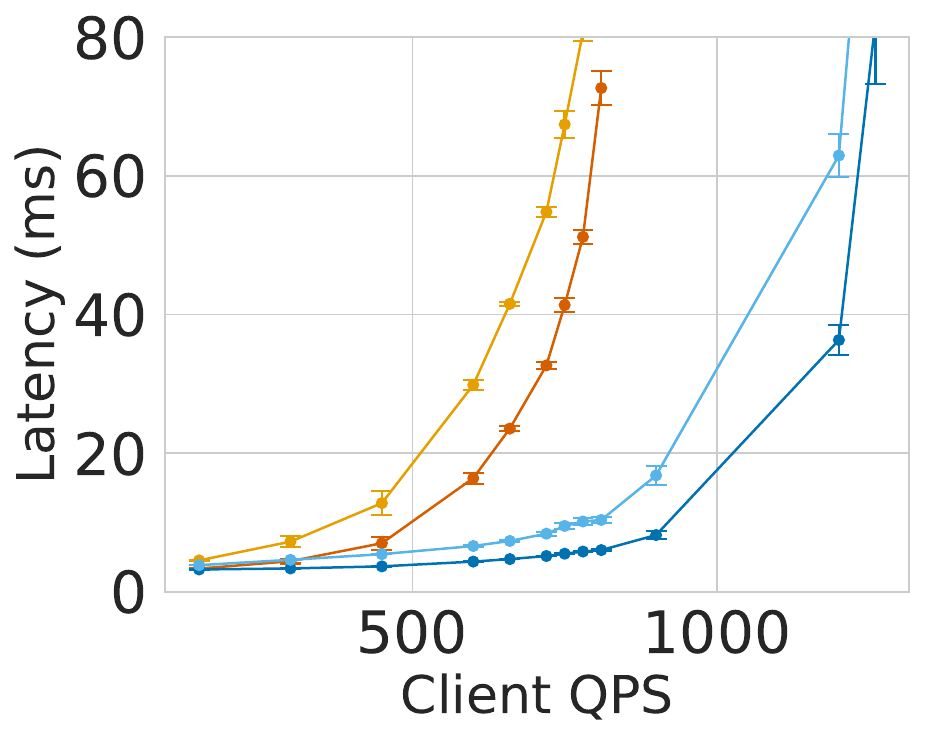}
        \caption{xapian++}
        \label{fig:xapian++lvs}
    \end{subfigure}
    \hfill
    \begin{subfigure}{0.22\linewidth}
        \includegraphics[trim=30 0 0 0,clip,width=\linewidth]{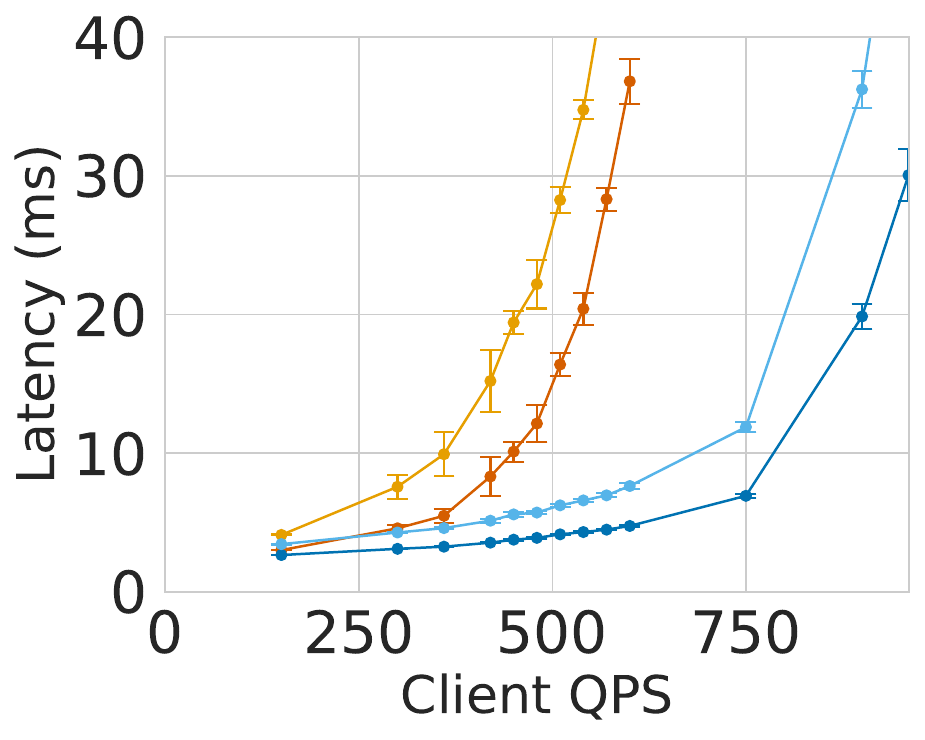}
        \caption{moses++}
        \label{fig:moses++lvs}
    \end{subfigure}

    \caption{Single- vs. multi-server characterization of TailBench++ applications.}
    \label{fig:singleVSMultiple++}
\end{figure*}

TailBench++ expands the scope of studies covered by the original TailBench since the introduced features enable reproducing dynamic scenarios involving interactions between multiple clients and multiple servers.

To illustrate its capabilities, this section examines the behavior of TailBench++ in a scenario involving a \textit{load balancer} that dynamically allocates clients across multiple servers. 
More specifically, three client connections are distributed among two servers using LVS.
As the objective is not to assess the performance of the load-balancing policy, the default round-robin algorithm is used.

Figure~\ref{fig:singleVSMultiple++} shows the results of the experiments 
for 95$^{th}$ and 99$^{th}$ latency percentiles for single-sever and multi-server scenarios. The X-axis and Y-axis represent the QPS issued by the clients and latency, respectively.
Given the natural variability in real systems --especially at high latencies-- the results include 95\% confidence intervals, depicted as error bars. These intervals reflect the variability across thirteen executions of the same experiment.
As expected, the multi-server scenario experiences lower latencies than the single-server scenario, with the exception of \texttt{specjbb++} and \texttt{silo++}, which do not benefit from the additional server. This observation is consistent with that observed in recent research work \cite{PONS2022194}. 
On the other hand, the latency variability remains similar between the single- and multi-server scenarios, as indicated by the comparable height of the error bars. This means that, as expected, the new features introduced in TailBench++ do not introduce additional variability.

\section{Example of Use Cases with TailBench++}
\label{sec:testing-new-features}

In this section, we show three case studies illustrating the new features of TailBench++. 
For illustrative purposes, the case studies use the \texttt{xapian} application. All the experiments analyze the tail latency (95$^{th}$ and/or 99$^{th}$ percentile) varying the QPS. For comparison purposes, some experiments also include the resulting mean latency.

\subsection{Interleaved Client Arrival Pattern}

\begin{figure}[tb]
    \centering
    \includegraphics[width=0.6\linewidth]{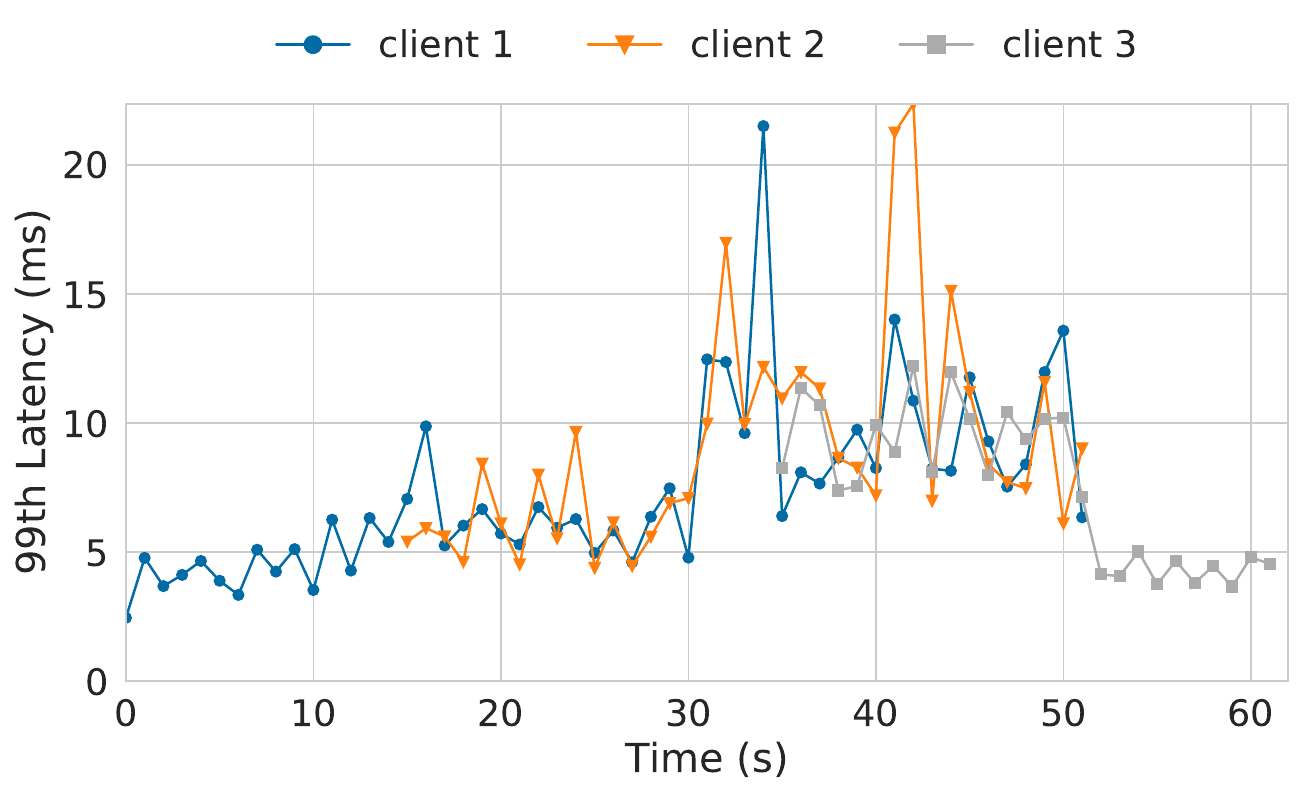}
    \caption{99$^{th}$ latency results for each interval, comparing three clients with the same QPS, but different starting times and total request counts.}
    \label{fig:clientDecide}
\end{figure}

This first case study illustrates the first three features: \textit{Feature 1. Unconstrained number of clients}, \textit{Feature 2. Persistent server}, and \textit{Feature 3. Independent client behavior}. 
For this purpose, the \texttt{xapian} benchmark is launched with one server instance, and three client processes (\textit{Clients 1, 2,} and \textit{3}), all of them with the same QPS rate (200) but each with a different starting time (seconds 0, 15, and 35) and a different total number of requests value (10000, 7000, and 5000). This scenario gives a different time window to each client: \emph{Client 1} runs for 50 seconds, \emph{Client 2} for 35 seconds, and \emph{Client 3} for 25 seconds. 
With the new persistent server functionality, TailBench++ can now accept connections from an unconstrained number of clients within a single experiment. Unlike TailBench, where the server must be aware of the exact number of clients beforehand and all clients must start at the same time, TailBench++ allows for more flexible and dynamic workload testing, thus removing these limitations.

Figure \ref{fig:clientDecide} shows the 99$^{th}$ tail latency of each client session during the execution. 
As the number of clients sending requests increases, the latencies perceived by each client also rise since the server must process a higher request rate. Notice that when \textit{Clients 1} and \textit{2} finish (second 50), the latency of \textit{Client 3} drops to values similar to those obtained by \textit{Client 1} when it was running alone at the start of the execution, which is reasonable as both generate the same QPS rate. 

\vspace{-0.2cm}
\subsection{Dynamic Client Load Pattern}

\begin{table}[tb]
    \vspace{-0.4cm}
    \caption{Order in which the client varies the QPS.}
    \vspace{0.2cm}
    \centering
    \begin{tabular}{|c|c|c|c|c|c|c|}
    \hline
        \textbf{Time interval (s)} & 0-9 & 10-19 & 20-29 & 30-39 & 40-49 & 50-59 \\
        \hline
        \textbf{QPS} & 100 & 300 & 500 & 600 & 800 & 100 \\
        \hline
    \end{tabular}
    
    \label{tab:orderqpsvariation}
\end{table}

\begin{figure}[tb]
    \centering
    \includegraphics[width=0.55\linewidth]{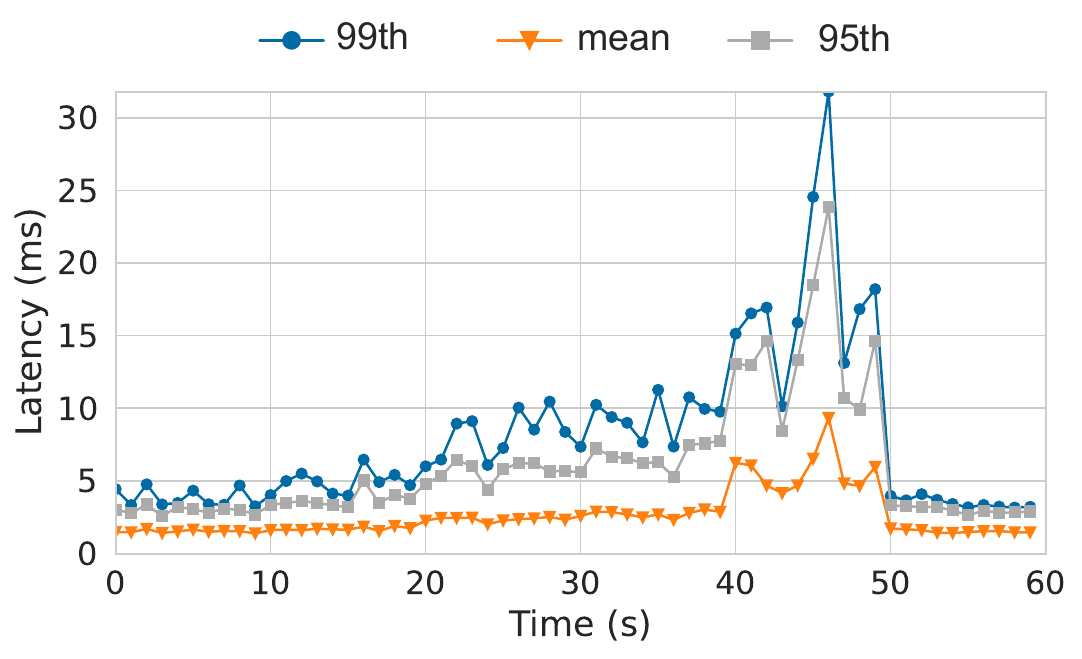}
    \caption{Latency results for one client process varying the QPS (see Table \ref{tab:orderqpsvariation}).}
    \label{fig:clientVariaQPS}
\end{figure}

The second case study illustrates \textit{Feature 4. Variable client load}. In this example, \texttt{xapian} is launched with one server instance and one client process. The client is set to change its QPS rate every 10-sec. time interval according to Table \ref{tab:orderqpsvariation}.
The client starts the execution (second 0) with 100 QPS; ten seconds later (second 10), the client increases the QPS to 300; the following two 10-second intervals, the client increases the QPS to 600 and 800, respectively. Finally, in the last 10-second interval (from the second 50 to 60), the client load drops to 100 QPS, the same load as at the beginning of the execution.

Figure \ref{fig:clientVariaQPS} shows how the latency varies for the client process during the execution.
The results show that the latency increases with the QPS. The highest latency values are found between seconds 40 and 50, where the client is exerting 800 QPS (the highest load). During this period, latency --particularly at the 95th and 99th percentiles-- exhibits a more bursty behavior, indicating that the server is nearing saturation. This means that a subset of the slowest requests is experiencing significantly higher delays.
Finally, notice that, as the QPS rate of the first and last execution intervals is the same, the latency in these two periods is similar.

\vspace{-0.2cm}
\subsection{Load Balancing Clients Across Multiple Servers}

\begin{figure}[tb]
    \centering
    \includegraphics[trim=0 330 0 0,clip,width=0.52\textwidth]{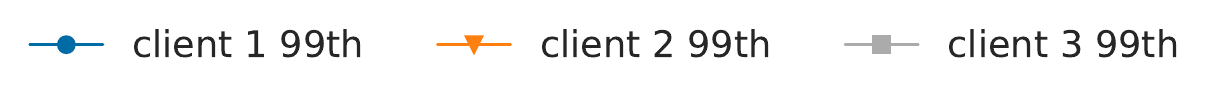}\\
      \captionsetup[subfloat]{position=bottom}
    \begin{subfigure}{0.495\linewidth}
        \includegraphics[trim=0 0 0 40,clip, width=\linewidth]{exp4base.pdf}
        \caption{Round-robin policy}
        \label{fig:base}
    \end{subfigure}
    \begin{subfigure}{0.47\linewidth}
        \includegraphics[trim=36 0 0 40,clip, width=\linewidth]{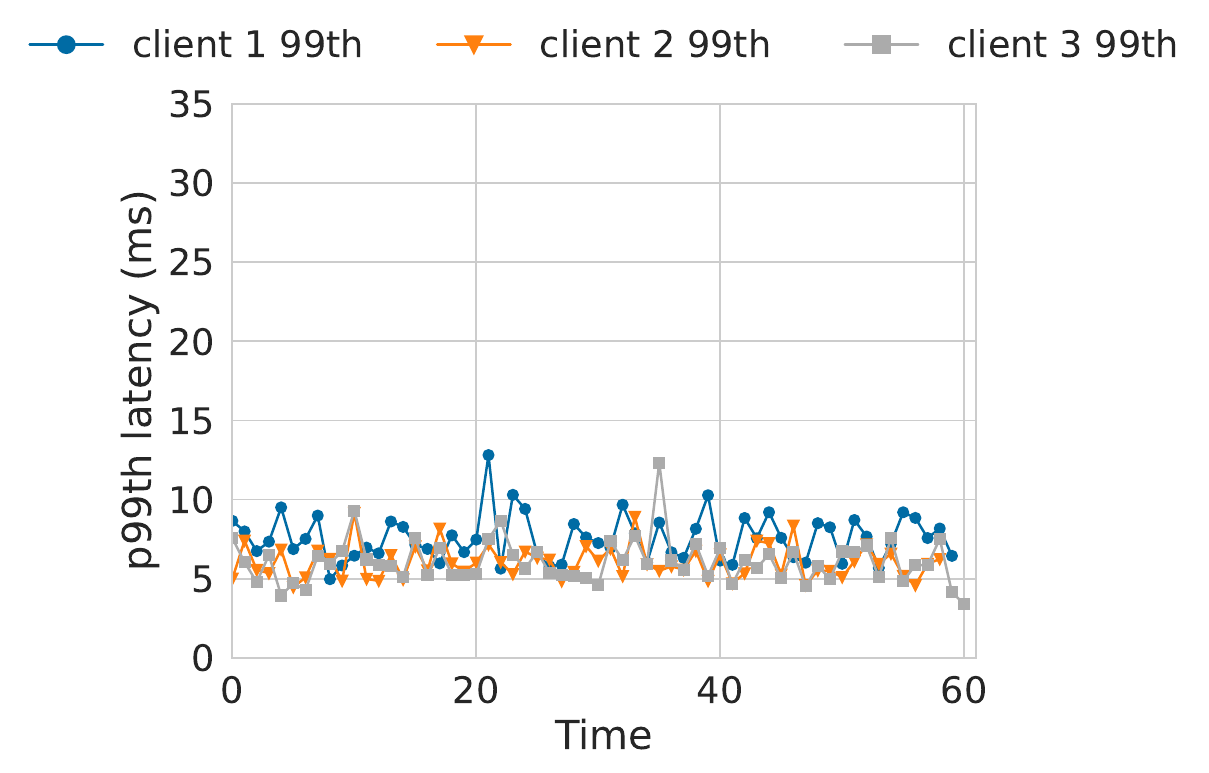}
        \caption{Load-aware policy}
        \label{fig:op}
    \end{subfigure}
    \caption{Comparison of the 99$^{th}$ percentile latency obtained by the clients under different load balancing policies.}
    \label{fig:load_balancing}
\end{figure}

This case study evaluates the capability of TailBench++ for being used in more complex deployment scenarios with multiple clients and servers. For this purpose, the \texttt{xapian} benchmark is launched with two servers and three clients which start running at the same time but with different request rates. \textit{Client 1} has a request rate of 500 queries per second (QPS), whereas \textit{Clients 2} and \textit{3} run with 200 QPS. To distribute client processes among servers, we use LVS under two different policies:  round-robin, which is widely used in cloud deployments, and a load-aware policy that aims to balance the request rate among servers. 

Figure \ref{fig:load_balancing} shows the results. The \textit{load-aware balancing policy} (right plot) distributes the connections so that the two clients with lower QPS (\textit{2} and \textit{3}) connect to the same server, leaving the remaining server dedicated to the client with higher QPS (\textit{1}). Meanwhile, round-robin merely distributes client connections among servers based on their arrival order at the director, resulting in a worse latency for \textit{Client 1} as it is assigned to the same server as \textit{Client 2}.

\vspace{-0.1cm}
\section{Conclusions}
\vspace{-0.1cm}

This paper has highlighted the lack of sufficient latency-critical benchmarks for cloud evaluation studies.
While TailBench, was identified as the most suitable suite for assessing the performance of cloud systems running latency-critical workloads, it cannot reproduce realistic, dynamic multi-client and multi-server environments.
To address this gap, we have proposed TailBench++, a benchmark suite designed to overcome the limitations of the original TailBench. We first analyzed existing benchmark suites, identified the limitations of TailBench, and developed TailBench++ as a more versatile tool for evaluating tail latency in complex cloud systems.
Through three case studies, we prove the effectiveness of TailBench++,  
showcasing the newly implemented features. 
By enabling dynamic and more realistic workloads, TailBench++ significantly expands the scope of cloud performance evaluation studies, especially for latency-critical applications. 

Overall, TailBench++ is aimed at helping researchers evaluate cloud systems, offering researchers a comprehensive and flexible framework to study tail latency in large-scale, dynamic environments. 
TailBench++ is publicly available at \url{https://github.com/zliUPV/Tailbenchplusplus}.

\begin{credits}
\subsubsection{\ackname} This work has been supported by the Spanish Ministerio de Ciencia e Innovaci\'{o}n and European ERDF under grants PID2021-123627OB-C51 and TED2021-130233B-C32. 

\subsubsection{\discintname} The authors have no competing interests to declare that are
relevant to the content of this article.
\end{credits}

\bibliographystyle{splncs04}
\bibliography{main}

\end{document}